\documentclass[prl,twocolumn,showpacs]{revtex4}

\usepackage{amsmath}
\usepackage{amsfonts}
\usepackage{amsthm}
\usepackage{amssymb}
\usepackage{amscd}
\usepackage[british]{babel}
\usepackage{graphicx}
\usepackage{psfrag}
\usepackage{epsfig}
\usepackage{rotating}
\usepackage{times}

\theoremstyle{plain}

\newcommand{\boxend}{\flushright{$\Box$}}

\newcommand{\w}{\omega}

\renewcommand{\tilde}{\widetilde}

\begin{document}

\title{Gravitational particle production in massive \\ chaotic inflation and the moduli problem}

\author{Jaume de Haro$^{a,}$\footnote{E-mail: jaime.haro@upc.edu}  and Emilio Elizalde$^{b,}$\footnote{E-mail: elizalde@ieec.uab.es, elizalde@math.mit.edu}}

\affiliation{$^a$Departament de Matem\`atica Aplicada I, Universitat
Polit\`ecnica de Catalunya, Diagonal 647, 08028 Barcelona, Spain \\
$^b$Instituto de Ciencias del Espacio (CSIC) \& Institut
d'Estudis Espacials de Catalunya (IEEC/CSIC)\\ Campus UAB, Facultat
de Ci\`encies, Torre C5-Parell-2a planta, 08193 Bellaterra
(Barcelona) Spain}

\thispagestyle{empty}

\begin{abstract}
Particle production from vacuum fluctuations during inflation is briefly revisited.
The moduli problem occurring with light particles produced at the end of inflation
is addressed, namely the fact that some results are in disagreement with
nucleosynthesis constrains.
A universal solution to this problem is found which leads to reasonable reheating
temperatures in all cases. It invokes the assumption that, immediately after
inflation, the moduli evolve like non-relativistic matter. The assumption is
justified in the context of massive chaotic inflation were, at the end of
inflation, the universe evolves as if it was matter-dominated.
\end{abstract}


\maketitle

{\it 1. Introduction.}---Particle production from vacuum fluctuations during inflation has been considered for some time (see, for instance, \cite{ckr98, kt99}). However, for light particles the results obtained have been reported not to be in agreement with constrains coming from nucleosynthesis. This is the so called {\it moduli problem} related to cosmological particle production.
In this letter we will try to better understand this issue in an attempt to clarify the
existing conflict between the results obtained in \cite{fkl99} and \cite{fmsvv09} concerning such moduli problem. Being more precise, for a light moduli field, in Ref.~\cite{fkl99} an abnormally small reheating temperature was obtained while, quite on the contrary, in Ref.~\cite{fmsvv09} an abnormally big one was reported.

We will here try to point out the basic differences between the approaches used in each of these papers and then propose a common solution to the moduli problem, in the context of massive chaotic inflation. To summarize the situation, it is our view that a common main assumption made in each one of the above references was to establish that, after inflation, the magnitude of the moduli field remains almost constant for a while, until it starts to oscillate. Assuming, on the contrary, that the moduli field actually evolves, after inflation, like non-relativistic matter, the problem is most naturally avoided in both cases. This is quite remarkable given the fact that the previous results where so extremely different in the two cases considered. To justify the new assumption, in the context of massive chaotic inflation one may argue that at the end of inflation the universe evolves in fact like a matter-dominated one. This will be explained in detailed in what follows.

Using the same notation as in Ref.~\cite{fmsvv09}, the  massive inflaton field, with mass $m$, will be called $\phi$, and the moduli fields $\chi$, the corresponding mass being $m_{\chi}$. We will also set $\alpha=m_{\chi}^2/m^2$  (notice that in \cite{fkl99} the mass of the inflaton is denoted by $M$ and the one of the moduli, by $m$).
\medskip

{\it 2. The massive inflationary model.}---In this model, the Friedmann equation is given by
\begin{eqnarray}\label{a1}
 H^2=\frac{\rho_{\phi}}{3M_p^2}=\frac{1}{6M_p^2}\left(\dot{\phi}^2+2V\right),
\end{eqnarray}
where $V=m^2\phi^2/2$, $M_p\sim 10^{18}$ GeV is the reduced Planck mass, and the inflaton field equation reads
\begin{eqnarray}\label{a2}
 \ddot{\phi}+3H\dot{\phi}+m^2\phi=0.
\end{eqnarray}
In the slow-roll approximation, that is, when the following equations hold,
\begin{eqnarray}\label{a3}
 H^2=\frac{m^2\phi^2}{6M_p^2},\qquad 3H\dot{\phi}+m^2\phi=0,
\end{eqnarray}
the evolution of the Hubble parameter is given by
\begin{eqnarray}\label{a4}
&& H(t)\cong H_0-\frac{m^2}{3}t,\qquad \phi(t)\cong \phi_0-\sqrt{2/3}mM_p t,\nonumber \\ &&
a(t)\cong e^{\frac{3}{2m^2}(H_0^2-H^2)}.
\end{eqnarray}
Inflation ends here at time $t_f\cong 3H_0/m^2$.  To obtain the approximate value of the Hubble parameter at the end of inflation, we impose that in this epoch the slow-roll parameters, namely
\begin{eqnarray}\label{a5}
\epsilon=\frac{M_p^2}{16\pi}\left(\frac{V_{\phi}}{V}\right)^2,\quad \eta=\frac{M_p^2}{8\pi}\left(\frac{V_{\phi\phi}}{V}\right),
\end{eqnarray}
be of order one. Then,  one has $\phi(t_f)\sim M_p$, which implies that $H(t_f)\sim m$.
Once the inflaton field strength drops below the corresponding Planck mass value, it begins to oscillate and the universe expands, like in the case of a matter-dominated universe (see, e.g.,  \cite{m05}, pgs.~235-43).

On the other hand, assuming that the moduli field is minimally coupled to gravity, then the equation for the inhomogeneous Fourier modes reads
\begin{eqnarray}\label{a6}
\ddot{\chi}_k+3H\dot{\chi}_k+\left(\frac{k^2}{a^2}+m^2_{\chi}\right)\chi_k=0.
\end{eqnarray}
Introducing now the variable $\tilde{\chi}=a\chi$, and using conformal time, this equation can be written as
\begin{eqnarray}\label{a7}{\tilde{\chi}}''_k+\left[{k^2}+{a^2}\left(m^2_{\chi}
-\frac{a''}{a^3}\right)\right]\tilde{\chi}_k=0.
\end{eqnarray}
The energy density, $\rho_{\chi}$, is given by
\begin{eqnarray}\label{a8}
\rho_{\chi}&=&\frac{1}{4\pi^2}\int dk \left[|\dot{\chi}_k|^2+\left(\frac{k^2}{a^2}+m^2_{\chi}\right)|\chi_k|^2\right] \nonumber \\ &=&
\frac{1}{2}\left(\langle\dot{\chi}^2\rangle+\frac{1}{a^2}\langle(\nabla\chi)^2\rangle+
m^2_{\chi}\langle\chi^2\rangle\right).
\end{eqnarray}
This quantity is divergent and needs to be re-normalized. One way to do that is to use adiabatic substraction, which cancels the ultraviolet modes and thus in fact only infrared modes contribute to the energy density. Actually, we are only interested in modes which
leave the horizon during inflation. Thus, at the end of the inflation era we only consider modes satisfying $H_0<k<\epsilon H(t_f)a(t_f)$, where $\epsilon$ is a dimensionless parameter smaller than one. This guaranties that those modes are well outside the horizon
at the end of inflation. Choosing $\epsilon\leq \alpha$ , the term $\frac{1}{a^2}\langle(\nabla\chi)^2\rangle$ turns out to be sub-leading.
Moreover, if the moduli field is light (i.e., $\alpha\ll 1$)---which is the case of interest, at the end of inflation---Eq.~(\ref{a7}) becomes
\begin{eqnarray}\label{a9}{\tilde{\chi}}''_k-\frac{a''}{a}\tilde{\chi}_k=0,
\end{eqnarray}
and has as solution $\tilde{\chi}_k=B_k a$, where $B_k$ is a constant. This means that ${\chi}_k$ is constant at the end of inflation. Thus, the term $\langle\dot{\chi}^2\rangle$ is also sub-leading. As a consequence, in this situation one has $\rho_{\chi}\sim m^2_{\chi}\langle\chi^2\rangle/2$.

At this point, in order to calculate the quantity $\langle\chi^2\rangle$ during inflation,
we first consider the de Sitter phase, in which case if we only take into account modes that leave the Hubble horizon, we get
\cite{li05}
\begin{eqnarray}\label{a10}
 \langle\chi^2\rangle\cong
\frac{3H_0^4}{8m_{\chi}^2\pi^2}\left(
1-e^{-\frac{2m_{\chi}^2t}{3H_0}}\right),
\end{eqnarray}
and taking now the derivative which respect to the cosmic time, we obtain
 \begin{eqnarray}\label{a11}
 \frac{d\langle \chi^2\rangle}{dt}+\frac{2m_{\chi}^2}{3H_0}\langle \chi^2\rangle=\frac{H^3_0}{4\pi^2}.
\end{eqnarray}
This can be generalized in terms of the stochastic equation \cite{fmsvv09}
\begin{eqnarray}\label{a12}
 \frac{d\langle \chi^2\rangle}{dt}+\frac{2m_{\chi}^2}{3H(t)}\langle \chi^2\rangle=\frac{H^3(t)}{4\pi^2},
\end{eqnarray}
which solution, for massive chaotic inflation, can be explicitly calculated:
\begin{eqnarray}\label{a13}
 {\langle \chi^2\rangle}=\frac{3H^{2\alpha}(t)(H^{4-2\alpha}_0-H^{4-2\alpha}(t))}{8(2-\alpha)\pi^2m^2}.
\end{eqnarray}
Here we have assumed that, at the beginning of inflation, $\chi$ is not a condensate (i.e., ${\langle \chi^2\rangle}=0$, initially).

For light particles, we have $\alpha\ll 1$ and, at the end of inflation, we get
\begin{eqnarray}\label{a14}
 {\langle \chi^2\rangle}=\frac{3H^4_0}{16\pi^2m^2},
\end{eqnarray}
which is much smaller than the result obtained in the pure de Sitter case \cite{bd78}
\begin{eqnarray}\label{a15}
 {\langle \chi^2\rangle}=\frac{3H^4_0}{8\pi^2m_{\chi}^2},
\end{eqnarray}
because $m_{\chi}\ll m$.
\medskip

{\it 3. The moduli problem arising from gravitational particle production.}---As we have seen, the gravitational particle production rate in the inflationary epoch could be very high, allowing for the possibility of creating a condensate (moduli field). Since at the end of
inflation this field behaves as non-relativistic matter, it eventually dominates the energy content of the universe, thus violating standard nucleosynthesis predictions, what will happen unless the amplitude of the moduli field is sufficiently small. The most stringent bound for successful nucleosynthesis prediction requires---provided the moduli field never dominates the expansion of the universe---that the ratio of the number density of light moduli produced during inflation, namely $n_{\chi}$, to the entropy of the universe after reheating, namely $s$, must satisfy the following inequality \cite{rt94,ak99}:
\begin{eqnarray}\label{a16}\frac{n_{\chi}}{s}\lesssim 10^{-12}-10^{-15},\end{eqnarray}
for moduli masses of the order of $10^2-10^3$ GeV.

In Ref.~\cite{fkl99}, the following expression was obtained:
\begin{eqnarray}\label{a17}\frac{n_{\chi}}{s}\sim  \frac{T_R \langle
\chi^2\rangle}{3m_{\chi}M_p^2}=\frac{H_0^4T_R}{16\pi^2m^2m_{\chi}M_p^2},\end{eqnarray}
where $T_R$ is the reheating temperature,
and $H_0$ the value of the Hubble parameter at the beginning
of inflation. (This formula comes from Eqs.~$(7)$ and $(33)$ of Ref.~\cite{fmsvv09}).
On the other hand, in Ref.~\cite{fmsvv09}, for light moduli fields, the result was:
\begin{eqnarray}\label{a18}\frac{n_{\chi}}{s}\sim \frac{3\alpha T_R H^{-2}
(t_R)H_0^{4}}{128\pi^2m_{\chi}M_p^2}\varpropto T_{R}^{-3},\end{eqnarray}
where  $t_R$ is the time at which the reheating ends.

If we consider the following realistic values: $m\sim 10^{-6}M_p$ (which ensures that the model does not overproduce anisotropies in the CMB signal), $H_0\sim 10^{-4}M_p$ (to obtain $10^4$ e-folds), and $m_{\chi}\sim 10^{-16}M_p$ (since we assume that the light moduli has a mass in the range of the gravitino one), then Eq.~(\ref{a17}) yields the constraint $T_R\lesssim 10^{-4}$ GeV, which is indeed an abnormally small reheating temperature. On the other hand, using the relation
$3H^2(t_R)M_p^2=\frac{\pi^2 gT_R^4}{30}$, where $g$ is the number of light degrees of freedom (for instance, $g=106.75$ for the standard model), Eq.~(\ref{a18}) yields $T_R \gtrsim 10^{16}$ GeV, which is in contradiction with the strong requirement that the grand unified theory symmetries are not restored---in order to prevent a second stage of monopole production (see, for example, \cite{btw}).

With the aim to understand these results we will now review the derivation of Eqs.~(\ref{a17}) and (\ref{a18}). To obtain the first one, the authors assume that reheating occurs after the beginning of the oscillations of the moduli field. This gives the following bound to the reheating temperature:
$T_R<\sqrt{m_{\chi}M_p}\sim 10^{9}$ GeV, which is needed to avoid the gravitino problem \cite{fkl99,btw}. They also assume that the field $\chi$ does not oscillate and that it (almost) does not change its magnitude up to the time, $\bar{t}$, when $H^2(\bar{t})$ becomes smaller than $m_{\chi}^2/3$.
There, the ratio of the energy density of the moduli field with respect to the energy density of the inflaton field, at time $\bar{t}$, is
\begin{eqnarray}\label{a19}
 \frac{\rho_{\chi}}{\rho_{\phi}}\sim \frac{m_{\chi}^2\langle\chi^2\rangle}{6H^2(\bar{t})M_p^2}\sim
\frac{\langle\chi^2\rangle}{2M_p^2},
\end{eqnarray}
where $\langle\chi^2\rangle$ is calculated at the end of inflation, since it is assumed that the amplitude of the moduli field almost does not change along the time interval $(t_f,\bar{t})$.

From the end of inflation to the reheating time our patch of universe evolves as if it were matter dominated and, thus, the ratio (\ref{a15}) is constant up to the reheating time. Then, as a consequence of the facts that at reheating time one has $\rho_{\phi}(t_R)=\frac{\pi^2 gT_R^4}{30}$, that the entropy of the produced particles is $s=\frac{2\pi^2 gT_R^3}{45}$, and that the number density of light moduli produced during inflation is given by $n_{\chi}\sim \frac{\rho_{\chi}}{m_{\chi}}$,  the expression (7) in Ref.~\cite{fkl99} was obtained, namely
\begin{eqnarray}\label{a20}
\frac{n_{\chi}}{s}\sim  \frac{T_R \langle
\chi^2\rangle}{3m_{\chi}M_p^2}.
\end{eqnarray}
To get the value of $\langle \chi^2\rangle$ one can assume that the moduli field is nearly massless, and then use Eq.~(\ref{a14}), which yields
\begin{eqnarray}\label{a21}
 {\langle \chi^2\rangle}=\frac{3H^4_0}{16\pi^2m^2}.
\end{eqnarray}
Finally, inserting it into Eq.~(\ref{a20}), one gets Eq.~(\ref{a17}).

Within a different approach, in Ref.~\cite{fmsvv09} immediate thermalization was assumed for the inflaton and for the moduli fields after inflation, what leads to
\begin{eqnarray}\label{a22}
 \frac{n_{\chi}}{s}\sim \frac{45m_{\chi}\langle \chi^2\rangle}{2\pi^2 g T_R^3}.
\end{eqnarray}
The value of $\langle \chi^2\rangle$ for a light field at the end of reheating is taken in \cite{fmsvv09} from Eq.~(\ref{a14}), what leads to the final result
\begin{eqnarray}\label{a23}\frac{n_{\chi}}{s}\sim
\frac{270\alpha  H_0^{4}}{128\pi^4 m_{\chi} g T_R^3   }
,\end{eqnarray}
in coincidence with Eq.~(31) of \cite{fmsvv09}.
\medskip

{\it 4. Universal solution to the moduli problem.}---After careful investigation, we here identify a problem which roots are common to both situations---even if, in principle, the approaches look so different, even more the results obtained in the two cases. Indeed, in Ref.~\cite{fkl99} it is assumed that the amplitude of the moduli field does not change from the end stage of inflation to
the beginning of the oscillations. Similarly, in Ref.~\cite{fmsvv09} the assumption is done that the solution (\ref{a14}) still holds after inflation.
What we will here defend is that, immediately after inflation ($H(t_f)\sim m$),
when the universe behaves as if it was matter dominated, the moduli actually {\it evolve like non-relativistic matter.} Then, for $t\in [t_f,t_R]$, one has
\begin{eqnarray}\label{a24}\frac{\rho_{\chi}(t)}{\rho_{\phi}(t)}=
\frac{\rho_{\chi}(t_f)}{\rho_{\phi}(t_f)}\sim \frac{m_{\chi}^2\langle\chi^2\rangle}{6H^2(t_{f})M_p^2}\sim
\frac{\alpha\langle\chi^2\rangle}{6M_p^2},
\end{eqnarray}
what means that
\begin{eqnarray}\label{a25}
{\rho_{\chi}(t_R)}\sim
\frac{\alpha\langle\chi^2\rangle}{M_p^2}\frac{\pi^2 gT_R^4}{180}
\end{eqnarray}
and, consequently,
\begin{eqnarray}\label{a26}
 \frac{n_{\chi}}{s}\sim \frac{\alpha T_R \langle \chi^2\rangle}{8M_p^2m_{\chi} }.
\end{eqnarray}
Note that this expression is essentially the same as Eq.~(3.3) of \cite{rt94}, the crucial difference with Eq.~(\ref{a20}) being the factor $\alpha$ in the numerator. To evaluate $\langle \chi^2\rangle$ at the end of inflation we use Eq.~(\ref{a14}), and obtain
\begin{eqnarray}\label{a27}
 \frac{n_{\chi}}{s}\sim \frac{3\alpha H_0^{4} T_R}{128\pi^2m^2 m_{\chi}M_p^2},
\end{eqnarray}
which, for realistic values: $m\sim 10^{-6} M_p$, $H_0\sim 10^{-4} M_p$, and $m_{\chi}\sim 10^{-16} M_p$, yields the perfectly consistent result $T_R\lesssim 10^{16}$ GeV.

The main reason which leads us to assume that, immediately after inflation has ended, the moduli evolve like non-relativistic matter is that the solution to Eq.~(\ref{a6}), in the massless case, in a matter-dominated universe, is given by \cite{bd82,m07}
\begin{eqnarray}
\chi_k(\eta)=\frac{1}{a^{3/2}}(k\eta)^{3/2}\left[c_1H_{3/2}^{(1)}(k\eta)+c_2H_{3/2}^{(2)}(k\eta)\right].
\end{eqnarray}
Thus, for the infrared modes, one has
\begin{eqnarray}
\chi_k(\eta)\sim\frac{1}{a^{3/2}},
\end{eqnarray}
which means that $\langle\chi^2\rangle\sim a^{-3}$. As a consequence, for light moduli it
is most reasonable to assume the same kind of behavior in the matter-dominated phase.

Another way to understand this very important point, namely the fact that the moduli 
must evolve like non-relativistic matter, comes actually from the very requirements 
of massive chaotic inflation: the moduli field satisfies the conservation equation
\begin{eqnarray}\label{A}
\dot{\rho}_{\chi}=-3H(\rho_{\chi}+p_{\chi}),
\end{eqnarray}
being $p_{\chi}$ the pressure. Therefore, as it occurs that in massive chaotic
inflation the universe enters, at the end of the inflation epoch, into a
matter-dominated phase, one has that $H(t)\sim \frac{2}{3t}$ and thus the solution
to Eq.~(\ref{A}) is ${\rho}_{\chi}\sim a^{-3}$ and $p_{\chi}\sim 0$.
Note, as a consequence, that although the mass of the field is rather
small, it just cannot evolve like relativistic
matter since it must necessarily satisfy Eq.~(\ref{A}), while it turns out that
relativistic matter will only satisfy this equation when $H(t)\sim \frac{1}{2t}$
(the scale factor would need to go as $a(t)\sim t^{1/2}$, and this is clearly not 
the case).

 \medskip

To finish, a couple of remarks are in order.

1. If at the beginning of inflation one takes the energy density to be of order $M_p^4$---which is the initial value considered, e.g., in \cite{l85,l86}---one gets $H_0\sim M_p$ and Eq.~(\ref{a27}) yields the constraint $T_R\lesssim 1$ GeV, which is indeed a very small value for the reheating temperature. Thus, in order to obtain a coherent result, one is necessarily compelled to assume that, initially, the Hubble parameter is smaller than $M_p$.

2. Numerical studies have been carried out with the purpose to calculate gravitational particle production. For instance, in Ref.~\cite{gtr99} a diagonalization method was used to calculate the production of minimally coupled light particles. As explained by the authors of that paper, such calculation is technically very difficult, for very small masses. Actually, what we do believe it really happens is that the method employed there only works well for {\it massive} conformally coupled particles \cite{gmm94,zs72}. In fact, the Bogoliubov beta coefficient is given by
\begin{eqnarray}\label{a30}
|\beta_{k}(\eta)|^2 &=&
\frac{1}{\omega_k(\eta)}\left(\frac{1}{2}\left[|\tilde{\chi}'_{k}(\eta)|^2 +
\omega_k^2(\eta)|\tilde{\chi}_{k}(\eta)|^2\right]\right. \nonumber \\ &&  \hspace*{3cm} - \left. \frac{1}{2}\omega_k(\eta)\right),
\end{eqnarray}
where
\begin{eqnarray}
&& \label{a31}\tilde{\chi}_{k}(\eta)=a(\eta)\chi_{k}(\eta), \nonumber \\
 &&\w_{ k}^2(\eta)\equiv k^2+a^2(\eta)m_{\chi}^2+(6\xi-1)\frac{a''(\eta)}{a(\eta)}.
\end{eqnarray}
Note that, in the minimally coupled case $\xi=0$, one has
\begin{eqnarray}
\w_{ k}^2(\eta)\equiv k^2+a^2(\eta)(m_{\chi}^2-2H^2-\dot{H}).
\end{eqnarray}
Then, if the moduli is very light,
for infrared modes this frequency is not well defined. Actually, in the de Sitter phase and for modes well outside the Hubble horizon, one obtains
\begin{eqnarray}
\w_{ k}^2(\eta)&\equiv& k^2+a^2(\eta)(m_{\chi}^2-2H^2)  \nonumber \\
&\sim& k^2-2a^2(\eta)H^2<0.
\end{eqnarray}
And, since these modes turn out to be the ones which are most important in order to calculate the relevant quantities \cite{li05,v83}, we would finally conclude that the diagonalization method is not the most appropriate one to be used in this case. What is more, in order to compute the desired quantity, the authors calculate $\rho_{\chi}/m_{\chi}$ numerically. However, $\rho_{\chi}$ is known to be a divergent quantity, if the field is not conformally coupled (see, e.g., \cite{b80}). It
must necessarily be renormalized previously, what renders utmost difficult to develop a rigorous procedure in order to perform such calculation numerically.

\medskip

This investigation has been
supported in part by MICINN (Spain), projects MTM2011-27739-C04-01,
FIS2006-02842 and FIS2010-15640, by CPAN Consolider Ingenio Project,
and by AGAUR (Generalitat de Ca\-ta\-lu\-nya), contracts 2009SGR-345
and 2009SGR-994.

\end{document}